\documentstyle[prl,aps]{revtex} 

\begin{document}
\title{Proximity Induced Josephson-Quasiparticle Process in a Single Electron 
Transistor}

\author{A.\ J. Manninen, R.\ S. Poikolainen, T.\ V. Ryyn\"{a}nen, and J.\ P. 
Pekola}

\address{Department of Physics, University of Jyv\"{a}skyl\"{a}, P.O. Box 35, 
FIN-40351 Jyv\"{a}skyl\"{a}, Finland}

\maketitle

\begin{abstract}
We have performed the first experiments in a superconductor - normal metal - 
superconductor single electron transistor in which there is an extra 
superconducting strip partially overlapping the normal metal island in good metal-
to-metal contact. Superconducting proximity effect gives rise to current peaks at 
voltages below the quasiparticle threshold. We interpret these peaks in terms of 
the Josephson-quasiparticle process and discuss their connection with the proximity 
induced energy gap in the normal metal island.  
\end{abstract}
\vspace{0.5cm}
\pacs{74.50.+r, 73.23.Hk, 73.40.Gk}

Proximity effect, i.e., penetration of the order parameter of a superconductor (S) 
into a normal metal (N) through a good contact, has been known and studied for 
almost 40 years \cite{deGeProx}. It is again under an active investigation. 
Progress in nanolithographic techniques has made it possible to study the proximity 
effect on length scales smaller than the normal coherence length, $\xi_N(T)$, which 
determines the depth of penetration of superconducting order into the normal metal 
(for a recent review see Ref.\ \cite{EstRev}). Evaporated metals are in the dirty 
limit, where $\xi_N (T) = (\hbar D/2 \pi k_B T)^{1/2} > l_e$; here $D = v_F l_e /3$ 
is the electronic diffusion constant, $v_F$ the Fermi velocity, and $l_e$ the 
elastic mean free path in N. 

Transport properties due to the proximity effect are very nontrivial. For example, 
the resistance of a normal metal wire in contact with a single superconductor can 
either increase or decrease even at length scales much larger than $\xi_N (T)$ when 
the sample is cooled below $T_c$, the transition temperature of the superconductor 
\cite{PetraJETP}. At very low temperatures the resistance approaches the normal-
state value \cite{CharReent}. On the other hand, a normal wire in good contact to a 
superconductor at both ends can transport a supercurrent if the length of the wire 
is of the order of $\xi_N (T)$ or shorter \cite{CourtCritcur}. 

We have performed electrical transport measurements in a new kind of a proximity 
structure: a superconductor - normal metal - superconductor single electron 
transistor (SNS-SET), which has an extra superconducting strip across the normal 
metal island in good metal-to-metal contact (see Fig.\ 1). We call this structure 
an SPS-SET where P stands for proximity-induced superconductivity. It has been 
predicted that the proximity effect should induce a constant energy gap $\Delta'$ 
into the central island if it is not too long compared with $\xi_N (T)$. $\Delta'$ 
is expected to be of the order of Thouless energy $E_T = \hbar D/L^2$ as long as 
$E_T < \Delta$, the energy gap of the superconducting strip \cite{BelzDOS,GoluDOS} 
(see Fig.\ 1 for our definition of $L$). This is in contrast with the semi-infinite 
structure studied by Gu{\'e}ron {\it et al.} \cite{GuerDOS} - a long normal metal 
wire in good contact with a superconductor on one end. In that structure a 
pseudogap, i.e., a nonzero minimum in the density of states at the Fermi energy, 
was observed in the normal metal at submicron distances from the NS contact, in a 
reasonable quantitative agreement with theory \cite{BelzDOS,GoluDOS,GuerDOS}.

We fabricate the samples on a nitridised silicon substrate by multiple angle 
evaporation through a mask made of a PMMA / P(MMA-MAA) double layer resist and 
patterned using electron beam lithography. Copper is used as a normal metal and 
aluminium as a superconductor. In our typical sample (see Fig.\ 1), a 21-nm thin 
layer of Al is first evaporated from "left" at an angle of 23 degrees with respect 
to the normal of the substrate. After oxidising in pure O$_2$ at $p \simeq 0.5$ 
mbar for about 30 seconds, a 5-nm layer of Al is evaporated from "right" at an 
angle of 60 degrees; in that way, the unwanted shadow of the vertical line is 
deposited on the wall of the resist as in Ref.\ \cite{CourtCritcur} and is removed 
in the liftoff. Immediately after evaporating Al, a 27-nm layer of Cu and another 
26-nm layer of Al are evaporated from "right" using angles of 23 and 60 degrees, 
respectively. In this process, AlO$_x$ tunnel junctions are formed between the 100-
nm wide vertical Cu line and the two 100-nm wide Al lines on the left. On the other 
hand, there is a pure metal-to-metal contact between the Cu line and the 200-nm 
wide Al strip on the right, without a tunnel barrier in between. Note that in this 
process a layer of Cu is deposited between Al films in some parts of the horizontal 
lines (shaded regions in Fig.\ 1). In order to check that this Cu layer does not 
affect our results, we have fabricated some of the samples with a slightly 
different structure, in which Cu is evaporated from "top" (see Fig.\ 1) at an angle 
of 35 degrees, such that the Cu and Al lines are separated. Results are similar in 
both structures.

The resistivity of our evaporated Cu film is typically 1.7$\cdot$10$^{-8}$ 
$\Omega$m at $T=4.2$ K. This gives $l_e = 39$ nm and $D=0.020$ m$^2$/s, 
corresponding to the coherence length $\xi_N (T) = T^{-1/2} \cdot 0.16$ $\mu$m when 
$T$ is expressed in kelvins. 

Figure 2 shows an example of the measured low-temperature current-voltage (IV) 
characteristics of an SNS-SET (Fig.\ 2a) and an SPS-SET (Fig.\ 2b), which were 
fabricated simultaneously on the same chip a few microns apart. The only difference 
between the samples is the extra Al strip in the SPS-SET. The SPS-SET has $L=0.65$ 
$\mu$m (see Fig.\ 1), such that we would expect the proximity effect to induce an 
energy gap $\Delta '$ of about $E_T \simeq 30$ $\mu$eV in the central island. The 
measurement was made at $T = 70$ mK, such that $k_BT \ll E_T$, and $\xi_N (T) 
\simeq 0.6$ $\mu$m was only slightly smaller than $L$. 

The quasiparticle current can flow through a single electron transistor only when 
the bias voltage $V$ exceeds a threshold $V_T$, which varies between $V_{T,\min}$ 
and $V_{T,\max}$, depending periodically on the gate voltage $V_g$. Ideally, 
$V_{T,\max} = V_{T,\min} + 2 E_C$ where $E_C = e^2/2C_\Sigma$ is the single-
electron charging energy and $C_\Sigma$ is the total capacitance of the central 
island; for an SNS-SET $V_{T,\min}=2 \Delta /e$, and for an SS'S-SET consisting of 
two superconductors with energy gaps $\Delta$ and $\Delta'$, respectively, 
$V_{T,\min}=2 (\Delta+\Delta') /e$. In the samples of Fig.\ 2, $E_C$ was made 
rather small, less than 20 $\mu$eV, such that the $V_g$ dependence of the IV curves 
is very weak. Based on the estimates presented above, we expected that $V_{T,\min}$ 
of the SPS-SET would be larger than that of the SNS-SET by $2 \Delta' /e \simeq 
2E_T/e \simeq 60$ $\mu$V. However, the measured large-scale IV characteristics of 
both samples are very similar. We have not observed any systematic difference in 
$V_{T,\min}$ of simultaneously fabricated SNS-SET / SPS-SET pairs. The 
reproducibility of this result is about 20 $\mu$V; there are differences of this 
magnitude even in $V_{T,\min}$ of two similar SNS-SETs fabricated simultaneously in 
immediate neighbourhood of each other.

However, when the IV characteristics of the SNS-SET and the SPS-SET are studied on 
a more sensitive current scale, they look very different from each other: that of 
the SNS-SET is smooth, whereas the SPS-SET has a clear current peak at $V \simeq 
0.2$ mV. Similar peaks were observed in all our 12 SPS-SET samples in which the Cu 
line and the extra Al strip were in direct metal-to-metal contact. On the other 
hand, no current peaks were observed if there was a tunnel barrier with resistance 
$R_T \simeq 10$ k$\Omega$ between the Cu line and the Al strip. Thus, the new peak 
is clearly caused by the proximity-induced superconductivity in the normal metal.

In order to explain the origin of this peak, we note that the so-called Josephson-
quasiparticle (JQP) process gives rise to very similar features in superconducting 
(SSS) single electron transistors \cite{FultonJQP,DimaJQP,SchonJQP,Nakamura}. In 
that process, a Cooper pair first tunnels into (or from) the island through one 
junction, and then two quasiparticles tunnel separately through the other junction. 
Because Cooper pair tunnelling can only take place when the energy does not change 
in the process, the JQP cycle is possible only at resonance voltages given by
\begin{equation}\label{VJQP}
V = \frac{4E_C}{e} (n \pm \frac{Q_0}{e}),
\end{equation}
where $n$ is an integer which is related to the number of extra electrons in the 
central island of the transistor. In our asymmetric voltage bias, in which one 
terminal of the transistor is grounded and the bias voltage $V$ is applied in the 
other terminal, $Q_0 = C_g(V_g - V/2) + Q_b$, where $C_g$ is the gate capacitance 
and $Q_b$ is the background charge induced by charged impurities near the island 
\cite{MatchP}. 

In an SS'S-SET the basic JQP process is possible at bias voltages $(\Delta + 
\Delta'+ E_C) /e < V < (\Delta + \Delta'+ 3E_C) /e$. Thus, when $E_C \ll 
\max(\Delta, \Delta')$, the JQP current peak is located at $V \simeq (\Delta + 
\Delta') /e \simeq V_T/2$ independent of the gate voltage, as we have earlier 
observed in SSS-SETs. The current peak of the SPS-SET of Fig.\ 2b has this 
property.

When $E_C$ is not very small, Eq.\ (\ref{VJQP}), which is valid even when the 
energy gap in the island of the transistor is different from that in the terminals, 
gives a series of lines in the $V_g-V$ plane at which the JQP cycle is possible. 
When $V$ is ramped at a constant $V_g$ as in Fig.\ 2 or, alternatively, $V_g$ is 
ramped at a constant $V$, a current peak can be observed when one of those lines is 
crossed. The latter method was used in the measurements of Fig.\ 3a, which shows 
current $I$ as a function of $V_g$ at several bias points $V < 2 \Delta /e$ in one 
of our SPS-SET samples. We can see clear current peaks which move when $V$ is 
changed. The positions of the peaks in the $V_g-V$ plane are plotted in Fig.\ 3b, 
with lines corresponding to Eq.\ (\ref{VJQP}) with $e/C_g = 1.0$ mV, $Q_b/e = 0.05$ 
and $E_C = 48$ $\mu$eV. The good fit supports strongly the interpretation of these 
features in terms of the JQP process. 

The fitted $E_C$ is in agreement with the depth $\Delta G$ of the zero-voltage dip 
of the dynamic conductance in normal state. The value measured at $T = 4.2$ K and 
scaled with the asymptotic conductance $G_T = 1/R_T$ is $\Delta G/G_T = 0.062$. If 
the electromagnetic environment of the transistor is not taken into account, we 
would expect a somewhat smaller value $\Delta G/G_T = E_C / (3k_BT) = 0.044$ when 
$E_C = 48$ $\mu$eV \cite{CBT}. However, assuming a resistive environment with $R 
\simeq Z_0 = 377$ $\Omega$, the free space impedance, as in Ref.\ \cite{Joyez}, the 
expected value increases to $\Delta G/G_T \simeq 0.058$, which agrees well with the 
measured value. The agreement was good in our other samples, too. On the other 
hand, the gate voltage dependence of the threshold voltage of quasiparticle 
current, $V_T$, was much weaker than expected. For the sample of Fig.\ 3, the 
measured $V_{T,\max}-V_{T,\min} \simeq 20$ $\mu$eV instead of the expected $2 E_C / 
e \simeq 96$ $\mu$eV.

A somewhat naive way to estimate the expected height $\Delta I$ of the JQP peak in 
the SPS-SET is to use the result derived for the SSS-SET \cite{DimaJQP,Nakamura}:
\begin{equation}\label{height}
\Delta I \simeq \frac{2e^3}{\hbar^2} \frac{{E_J}^2R_T}{eV+E_C} =
\frac{\Delta'^2[K(\sqrt{1-\Delta'^2/\Delta^2})]^2}{2R_Te(eV+E_C)} ,
\end{equation}
where we have used the Ambegaokar-Baratoff (AB) value \cite{AB-Supercur} for the 
Josephson coupling constant,
\begin{equation}\label{AB}
E_J = \frac{\hbar \Delta'}{2R_Te^2} K(\sqrt{1-\Delta'^2/\Delta^2}), 
\end{equation}
between two superconductors having energy gaps $\Delta$ and $\Delta'$ ($< \Delta$) 
at $T = 0$. Here $K(x)$ is the complete elliptic integral of the first kind and 
$R_T$ is the tunnelling resistance of a single junction. If we now assume that the 
proximity effect induces an energy gap $\Delta' \simeq E_T$ in the central island 
of the SPS-SET transistor, Eq.\ (\ref{height}) predicts a JQP peak with height 
$\Delta I \simeq 1.4$ nA for the sample of Fig.\ 2b, for example. The measured 
value is more than two orders of magnitude smaller, $\Delta I \simeq 8$ pA, 
suggesting that $\Delta'$ is about an order of magnitude smaller than expected.  
The observed width of the peak, $\Delta V = 53$ $\mu$V, is in a reasonable 
agreement with the expected $\Delta V \simeq \hbar (eV+E_C)/(e^3R_T) = 39$ $\mu$V. 

In all of our samples the observed peak height was much smaller than predicted by 
Eq.\ (\ref{height}), by a factor between about 30 and 300. This is in agreement 
with not observing any difference in $V_{T,\min}$ between SNS and SPS transistors. 
Equation (\ref{height}) is based on the AB value of $E_J$, which often predicts 
supercurrents which are orders of magnitude larger than those experimentally 
observed in samples with submicron Josephson junctions \cite{EilMart}. Note also 
that Eq.\ (\ref{AB}) would predict for the sample of Fig.\ 2b the maximum 
"supercurrent" near $V = 0$ of $I_c^{\max} \simeq eE_J/\hbar \simeq 2.7$ nA, 
whereas the observed value is only about 1 pA. The observed supercurrents were very 
small in our other SPS-SETs, as well.

If the JQP peaks in the SPS-SET are caused by a proximity-induced energy gap 
proportional to $E_T$ in the central island, their height $\Delta I$ should depend 
strongly on the dimension $L$ of the sample: because $E_T \propto L^{-2}$, Eq.\ 
(\ref{height}) predicts that $\Delta I$ is roughly proportional to $L^{-4}$. We 
tested this prediction in two samples, each of which consisted of three separate 
SPS-SETs with $L = 0.50$, 0.90 and 1.15 $\mu$m fabricated simultaneously on the 
same chip. In sample 1, $d$ was varied in such a way that the tunnel junctions were 
always near the ends of the Cu island, whereas in sample 2, $d$ was kept constant. 
The results are shown in Fig.\ 4, in which the small-current IV characteristics of 
the transistors are shown with the gate voltage $V_g$ adjusted in such a way that 
the JQP peak was strongest; this corresponds to the situation in which two JQP 
lines in Fig.\ 3b cross. The measurements were made at $T \simeq 40$ mK, such that 
$\xi_N(T) \simeq 0.8$ $\mu$m. Current is multiplied by the resistance of a single 
tunnel junction, $R_T$, in order to remove the $1/R_T$ dependence of the peak 
heights (see Eq.\ (\ref{height})). 

In both samples 1 (Fig.\ 4a) and 2 (Fig.\ 4b) the scaled peak height $\Delta(R_TI)$ 
depends strongly on $L$. The insets of Fig.\ 4 show the measured $\Delta(R_TI)$ as 
a function of $L^{-4}$, together with the prediction based on Eq.\ (\ref{height}) 
but divided by 55. Even though the observed peaks are again almost two orders of 
magnitude smaller than predicted by Eq.\ (\ref{height}), the dependence of the peak 
height on $L$ is in good agreement with the expectation based on the existence of 
an energy gap $\Delta' \propto E_T$ in the Cu island. This is especially remarkable 
in sample 2, in which only the length of the island was varied but the distance 
between the measuring junctions and the extra Al strip of the island was kept 
constant. This suggests strongly that the proximity effect really induces in the Cu 
island a constant energy gap, whose size depends on the length of the island.

In conclusion, we have studied the charge transport properties of a new kind of a 
single electron transistor, which consists of superconducting leads, normal island 
and a superconducting strip evaporated in metal-to-metal contact across the normal 
island. We have observed new current peaks caused by the superconducting proximity 
effect in this structure. Based on the gate voltage dependence of the positions of 
the features, we interpret them as the Josephson-quasiparticle (JQP) peaks. The 
dependence of the JQP peak heights on the dimensions of the sample suggests the 
existence of a proximity-induced energy gap in the normal metal island of the 
transistor. This gap is, however, smaller than the order-of-magnitude theoretical 
estimate that we have used, and too small to modify the large-scale current-voltage 
characteristics on an observable level. 

We thank the Academy of Finland for financial support.

\vspace{1.5cm}

{\bf FIGURE CAPTIONS}

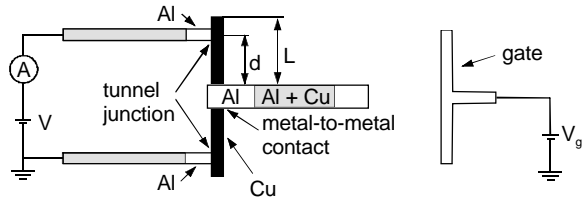
\begin{figure}  
\caption{The schematic diagram of the SPS single electron transistor. The Al strip, 
which is typically 3 $\mu$m long and 200 nm wide and which is in metal-to-metal 
contact across the 100 nm wide Cu island, gives rise to the proximity effect in the 
island. Colour code: white, Al (S); black, Cu (N); shaded, Al/Cu sandwich.}
\label{fig1}
\end{figure}

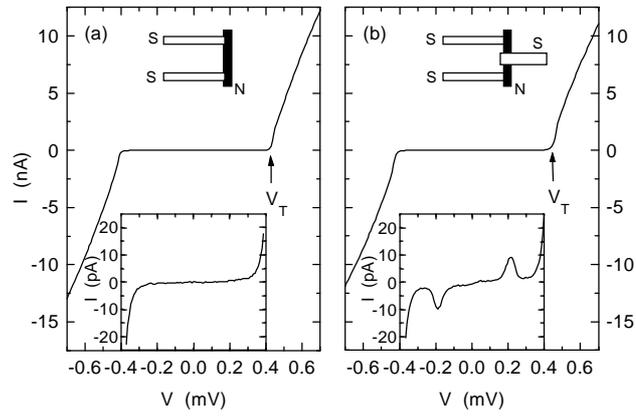
\begin{figure}  
\caption{Current-voltage characteristics of (a) an SNS-SET with $R_T = 22$ 
k$\Omega$ (the resistance of a single tunnel junction) and (b) an SPS-SET with $R_T 
= 21$ k$\Omega$, at $T = 70$ mK. In both samples $\Delta \simeq 220$ $\mu$eV and 
$E_C < 20$ $\mu$eV. In the SPS-SET $d = 0.40$ $\mu$m and $L = 0.65$ $\mu$m (see 
Fig.\ 1 for definition of $d$ and $L$); the SNS-SET has the same geometry except 
for the missing 200 nm wide Al strip across the island (see the upper insets). In 
large scale (main frames), no reproducible difference in the threshold voltage 
$V_T$ can be observed between the two transistors. In a more sensitive current 
scale (lower insets), the proximity effect gives rise to new current peaks in the 
SPS-SET.}
\label{fig2}
\end{figure}

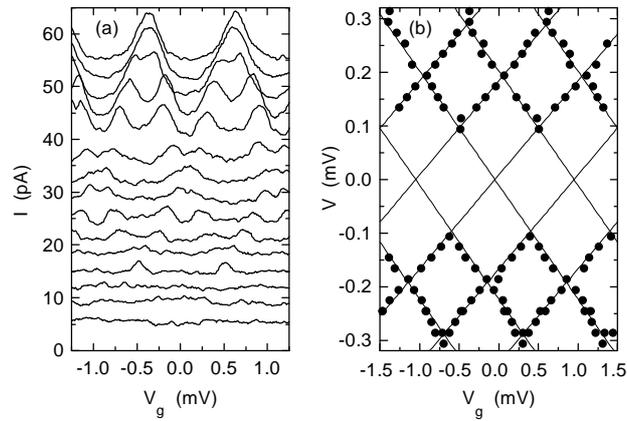
\begin{figure}  
\caption{(a) Current $I$ of an SPS-SET with $R_T = 17$ k$\Omega$, $d = 0.60$ 
$\mu$m, $L = 0.65$ $\mu$m and $\Delta \simeq 220$ $\mu$eV, as a function of gate 
voltage $V_g$ at bias points $V$ between 33 $\mu$V and 313 $\mu$V, with 20 $\mu$V 
intervals. Each trace is shifted up by 0.15 ($V/\mu$V) pA. (b) The locations of the 
observed current peaks in $V_g-V$ plane, together with lines corresponding to the 
expected JQP positions, Eq.\ (\ref{VJQP}), with $e/C_g = 1.0$ mV, $Q_b/e = 0.05$ 
and $E_C = 48$ $\mu$eV.}
\label{fig3}
\end{figure}

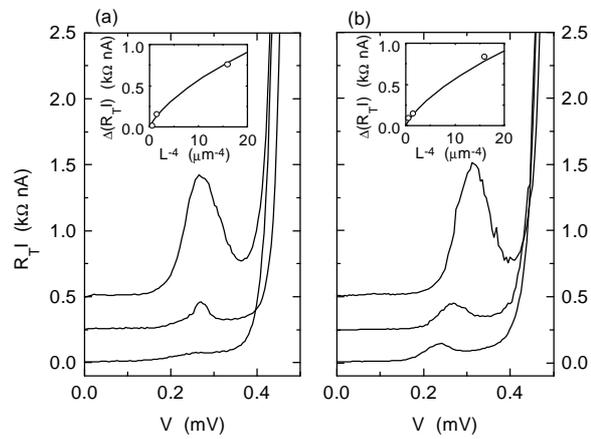
\begin{figure}  
\caption{Dependence of the IV characteristics of the SPS-SET on the length of the 
Cu island at $T = 40$ mK. In each trace $V_g$ has been adjusted in such a way that 
the JQP peak has its maximum height, and $I$ has been multiplied by $R_T$. The 
traces have been shifted along the vertical axis for clarity. In (a) the sample 
parameters are, from top to bottom: $L = 0.50$, 0.90, and 1.15 $\mu$m; $d = 0.45$, 
0.85, and 1.10 $\mu$m; $R_T = 29.6$, 32.0, and 35.2 k$\Omega$; $E_C = 70$, 65, and 
62 $\mu$eV.  Correspondingly in (b): $L = 0.50$, 0.90, and 1.15 $\mu$m; $d = 0.45$ 
$\mu$m for each SET; $R_T = 50.0$, 27.5, and 22.5 k$\Omega$; $E_C = 82$, 66, and 58 
$\mu$eV. Insets: the measured height of the current peak multiplied by $R_T$ as a 
function of $L^{-4}$ (open circles), together with the theoretical line based on 
Eq.\ (\ref{height}) but divided by 55. The values of the parameters used in the 
calculation are $D = 0.020$ m$^2$/s, $\Delta = 210$ $\mu$eV, $V = 250$ $\mu$V, and 
$E_C = 65$ $\mu$eV.}
\label{fig4}
\end{figure}

\end{document}